\newcommand{\squeezeup}{\vspace{-0.1mm}}

\documentclass[acmsmall, sigplan, 10pt, nonacm]{acmart}
\usepackage{microtype}

\AtBeginDocument{%
  \providecommand\BibTeX{{%
    \normalfont B\kern-0.5em{\scshape i\kern-0.25em b}\kern-0.8em\TeX}}}

\usepackage{xspace}

\setcopyright{rightsretained}
\acmPrice{}
\acmDOI{10.1145/3622758.3622894}
\acmYear{2023}
\copyrightyear{2023}
\acmSubmissionID{onward23papers-p95-p}
\acmISBN{979-8-4007-0388-1/23/10}
\acmConference[Onward! '23]{Proceedings of the 2023 ACM SIGPLAN International Symposium on New Ideas, New Paradigms, and Reflections on Programming and Software}{October 25--27, 2023}{Cascais, Portugal}
\acmBooktitle{Proceedings of the 2023 ACM SIGPLAN International Symposium on New Ideas, New Paradigms, and Reflections on Programming and Software (Onward! '23), October 25--27, 2023, Cascais, Portugal}
\received{2023-04-28}
\received[accepted]{2023-08-11}
\received[revised]{2023-10-12}

\begin{document}

\title{Concept-Centric Software Development}
\subtitle{An Experience Report}

\author{Peter Wilczynski}
\email{pwilczynski@palantir.com}
\orcid{0009-0000-1108-4447}
\affiliation{
    \institution{Palantir Technologies Inc.}
    \city{Denver}
    \state{CO}
    \country{USA}
}

\author{Taylor Gregoire-Wright}
\email{taylor@ontologize.com}
\orcid{0009-0008-6105-1801}
\affiliation{
    \institution{Ontologize LLC}
    \city{Palo Alto}
    \state{CA}
    \country{USA}
}

\author{Daniel Jackson}
\email{dnj@mit.edu}
\orcid{0000-0003-4864-078X}
\affiliation{
  \institution{Massachusetts Institute of Technology}
  \city{Cambridge}
  \state{MA}
  \country{USA}
}

\renewcommand{\shortauthors}{Wilczynski, Gregoire-Wright, and Jackson.}

\begin{abstract}
Developers have long recognized the importance of the concepts underlying the systems they build, and the primary role that concepts play in shaping user experience. To date, however, concepts have tended to be only implicit in software design with development being organized instead around more concrete artifacts (such as wireframes and code modules).

\palantir, a software company whose data analytics products are widely used by major corporations, recently reworked the internal representation of its software development process to bring concepts to the fore, making explicit the concepts underlying its products, including how they are clustered together, used in applications, and governed by teams. With a centralized repository of concepts, \palantir engineers are able to align products more closely based on shared concepts, evolve concepts in response to user needs, and communicate more effectively with non-engineering groups within the company.

This paper reports on \palantir's experiences to date, analyzing both successes and challenges, and offers advice to other organizations considering adopting a concept-centric approach to software development.


\end{abstract}








\newcommand{\palantir}{Palantir\xspace}
\newcommand{\foundry}{Foundry\xspace}
\newcommand{\gotham}{Gotham\xspace}
\newcommand{\apollo}{Apollo\xspace}

\maketitle

\section{Introduction} \label{introduction}

This paper reports on a major effort at \palantir, a leading producer of data analytics platforms used in many domains, to construct a concept inventory and integrate it into the company's development practices. 

This effort builds on a theory of concepts developed in a recent book, \textit{The Essence of Software} (henceforth EOS) \cite{jackson_essence_2021}, and outlined in Section \ref{eos_section} below. While EOS focuses on the detailed design of concepts, this paper addresses the larger concept development process---how concepts are invented, harmonized and refactored within a large development organization. This might be called “concept dynamics”, using terminology from business strategy that distinguishes “statics” (the study of game states) from “dynamics” (the study of game play itself) \cite{helmer_7_2016}.


Concepts are a vital substrate for cross-team collaboration. Because teams tend to be organized around ownership of functional subcomponents of the overall system, concepts often get stuck in the seams between separate development teams, and ossify and decay because no one team has the agency to make changes unilaterally. Small but subtle updates to existing concepts might allow them to generalize to new workflows, but lacking an understanding of the wider impact of these changes, teams are often reluctant to engage in this work, and choose instead to specialize locally, often reinventing the wheel. 

Concepts also play a role for the non-engineers in an organization, especially those in adjacent functions that require an understanding of the product such as sales, marketing and competitive analysis.




This project began at Palantir in 2022, and was led by the first two authors, the first a lead product manager and the second a former employee acting as a consultant. Given the primary role of ontologies in Palantir's own software, the strategy was to introduce concepts by augmenting Palantir's internal ontology with the concepts present in Palantir's software. 

This took the form of a concept repository built in Palantir's internal instance of its Foundry software platform. Beyond the core development functions, customer-facing solutions architects, product marketing, and legal teams all expressed interest in the concept repository. Much of the value of the project derived not just from making user-facing concepts explicit in Palantir's software, which enabled faster collaboration across product teams, but from the fact that by creating a concept object type in Palantir's internal ontology, concepts were automatically connected to the rest of Palantir's internal data. For example, a single concept was linked to the \textit{features} and \textit{applications} it was used by, the \textit{teams} responsible for building and maintaining those \textit{features} and \textit{applications}, the \textit{bug reports} and \textit{tickets} filed, company-wide product \textit{planning documentation}, and many more entities already present in the internal ontology.

Currently, the concept repository currently holds around 150 concepts, and is used by around 250 employees to refine, and align, the design of three major products.

The paper explains how the inventory was integrated into an existing document collection; gives examples of concepts and their evolution;  presents some of the successes and challenges that have emerged to date; and offers recommendations to others who might be interested in expanding the role of concepts in industrial software development.


\section{An Outline of Concept Design Theory} \label{eos_section}

For decades, software engineers have recognized the importance of the concepts that underlie software systems, but without a robust notion of what concepts are and how they might be described and analyzed. The theory of concept design described in \textit{The Essence of Software} \cite{jackson_essence_2021} (EOS) builds on well-established software engineering ideas, but also extends them in new directions. In this section, we outline some of the key ideas; for a fuller explanation readers are referred to the book \cite{jackson_essence_2021} and website (\hyperlink{https://essenceofsoftware.com}{https://essenceofsoftware.com}).

Concept design starts with the idea that many of the essential qualities of a software system (including its usability, robustness, maintainability, etc) follow from the functionality of the system and its structure. Concepts offer a way to define the functionality in terms of reusable, independent units that are well understood by users and aligned with their needs.

The idea that a compelling conceptual model is the key to usability goes back to early work on user-centered design \cite{norman-draper-ucd-book}; what is novel here is (a) the focus on explicit design of the conceptual model (rather than taking it as given and focusing instead on a faithful projection in the user interface); (b) decomposing the model into modular parts; and (c) identifying concepts not just in the data model (that is the structure of the abstract state) but also in the behavior (the actions that read and write the state).

The principles of concept design include: \textit{specificity}---that each concept should have one clearly defined purpose; \textit{familiarity}---that whenever possible a design should reuse a concept that is familiar to users from other applications; and \textit{independence}---that each concept should be defined without any dependence on other concepts, so that concepts (unlike features, which often rely on a context of some base functionality) can be grasped one at a time by users, and can be designed and analyzed separately. EOS offers a particular structure for defining concepts, and specifying them rigorously as state machines (with each concept having its own data model and actions).

The purported benefits of concept design include: achieving clarity and  simplicity in design; improving usability by encouraging  familiar, orthogonal and well motivated functions; aligning user experience across products in a family; and bridging the gaps between different roles so that UX designers, product managers, engineers, marketers, and so on, all share the same understanding.

In this project at Palantir, the emphasis has been on identifying the core concepts of existing products, and making explicit the processes by which they are developed, refined, extended and shared. While adopting the central ideas of EOS in identifying concepts as key assets and placing concept design at the center of development, in our efforts to see concepts widely used at the company we were reluctant to expect too much in terms of the exact form in which concepts are described. Moreover, many of the concepts that we identified are less mutually independent than EOS would demand. Nevertheless, as this paper reports, we have found that the introduction of a concept design perspective, with the explicit articulation of concepts as the elements of design, is already having some profound consequences.


\section{Conceptual Entropy and Complexity} \label{growth_of_complexity}
\begin{quote}
I will contend that Conceptual Integrity is the most important consideration in system design. It is better to have a system omit certain anomalous features and improvements, but to reflect one set of design ideas, than to have one that contains many good but independent and uncoordinated ideas.---Fred Brooks \cite{brooks_mythical_1995}
\end{quote}

\noindent \palantir has three main products---\foundry, \gotham and \apollo---developed by more than 1,000 software engineers working closely with product managers and designers. Like most other software development organizations, the company has seen a substantial increase in the complexity of its products and processes over time.


As of April, 2023 these products consist of ~130 discrete user-facing applications which in turn rely on ~2,800 discrete backend services. The source code for these components lives in ~1,700 individual Git repositories, the largest of which contains nearly 9M lines of code. In the first three months of 2023, ~1,300 contributors merged ~20k pull requests. Including ~480k automated contributions, the total diff during that period was ~54M lines of code.

Increases in complexity are inevitable as a company and its products grow in size and capability. We have come to believe, however, that although much of the complexity of  \palantir's products is an inevitable consequence of their power and flexibility, there are other respects in which complexity might be significantly reduced with positive impacts for both users and developers \cite{complexity_and_strategy}. Such complexity obscures important opportunities to improve our products. While we continue to increase the power and flexibility of our products dramatically, it has been harder to maintain clarity and consistency, both within and across products. As a result, users are less able to learn how to use our products and to exploit new features effectively. We have noticed in particular a growing class of bug reports that cannot be easily attributed to particular parts of the code, but which correspond instead to conceptual flaws and inconsistencies that cross module (and even product) boundaries.

The result is that what aspires to be a tightly integrated suite of products built around a core of common ideas slowly transforms into a collection of seemingly disconnected products that become increasingly fragmented over time. The underlying conceptual problems are often signaled by confusing terminology. As one user of our system noted: “\foundry's language has become more inconsistent. Similar features have different descriptions or just work differently overall. [Pipeline] Builder's ‘master' branch is called ‘main', and its ‘branches' are called ‘sandbox[es]'. As we move across different apps, it confuses us why features which are intended to do the same thing are named differently. It breaks our mental model and interrupts our workflow.”

\palantir is not alone in facing the challenge of aligning products across a diverse family. Regarding Google's products, technology reporter Casey Newton advised in a September 2022 tweet \cite{newton_2022}: “Google Reminders are now Google Tasks. To create a Reminder, use Calendar. To create a Task, use Gmail. Do NOT use Keep. Hope this helps.” The history of Google's messaging apps likewise documents fifteen years of unfettered expansion \cite{amadeo_2021}.

These problems are not due to a lack of sophistication in tooling. Almost all companies have systems, often highly customized, for source code management, continuous integration, continuous delivery, live upgrades, realtime observability, issue tracking, project management, and customer feedback. Most of these tools, however, support the addition of new features that \textit{expand} a product---and, indeed, we can now build products and grow them faster than ever before. But \textit{contractionary} pressure is somehow harder to muster and support. Even at Meta, a notably centralized product organization, harmonizing three messenger apps took years of sustained focus and commitment from thousands of people \cite{isaac_2019}, and this was within a category (chat) whose concepts are unusually stable and well-understood.

To address these problems, we believe attention must be redirected towards the concepts that underlie our products: the essential elements of functionality that characterize the products and that users need to understand in order to use them. In our diagnosis, inconsistencies within and across products are often the result of different concepts being used for the same functionality, and the difficulties of users in exploiting functionality arise from unclear or needlessly complicated concepts. Complexity outpaces functionality when concepts proliferate and are degraded by special cases and unnecessary couplings.

By analogy to physical systems that suffer a loss of structure over time, we call this phenomenon ``conceptual entropy.'' In its simplest form, conceptual entropy is the number of concept pairs that involve duplication or confusion: two concepts that share the same name but have different meanings, or that each share the same meaning but have different names. Loss of uniformity and clarity also contribute to conceptual entropy, as seemingly arbitrary accretions of functionality complicate and corrupt concept structure.

Just as the entropy of a physical system will tend to increase over time, so the conceptual entropy of a software system will increase too as new features are added, features are intentionally changed, refactorings introduce subtle differences to features, and organizations acquire other products and codebases. And just as reducing the entropy of a physical system takes work, the same is true with software---decreasing conceptual entropy is not free.

Technology trends impact growth in entropy. In our experience, microservice-oriented architectures may increase the growth rate if applications are overly coupled to a single backend service, while GraphQL can decrease the growth rate by making concepts spread across many microservices more accessible to individual applications.

But technology alone will not create the contractionary pressure necessary to develop large-scale integrated product suites. The only way to address entropy effectively, we believe, is to recognize it as a critical kind of technical debt, and to make sure that the evolution and extension of functionality is accompanied by a concerted effort to “condense” and clarify the underlying conceptual structures.

In \textit{Mythical Man Month} \cite{brooks_mythical_1995} (and later in his “Silver Bullet” paper \cite{brooks1987essence}), Brooks argued that the design of conceptual structures was central, and that strong concepts were the key to taming complexity. This observation rings true to anyone who has worked in large-scale software development. And yet, fifty years after Brooks first made this observation, there are few methods and tools for measuring and managing conceptual integrity. In this paper, we describe a tool we have deployed at \palantir to make concepts central in our development, to counter the increase of conceptual entropy, and to reorient us toward the holy grail of modern software development: achieving conceptual integrity at scale.

\section{An Example Concept} \label{an_example_concept}

To make the idea of concepts concrete, consider the \textit{Clip} concept (discussed in Section \ref{case_study_clips}). The need for the concept arose from a difficult dilemma. Users wanted to embed visualizations of data in presentations. Including a live link whose displayed content updates dynamically offers the advantage of attribution and consistency. But it also introduces new problems: when the referenced content is not accessible, no data would appear; and sometimes presentation authors want to snapshot the referenced content to prevent the risk of it changing unexpectedly during a high-stakes presentation.

The solution to this dilemma was to invent a new concept, called \textit{Clip}, that incorporates both options, allowing a presenter to embed a link and make flexible decisions later on whether the displayed content should correspond to the latest state or an earlier snapshot. This example illustrates several key distinguishing qualities of concepts \cite{jackson_essence_2021}:
\begin{itemize}
    \item \textbf{User-facing.} A concept represents functionality that is experienced directly by the user, and not a code module whose impact on the user can only be understood in the context of how that module is called. The \textit{Clip} concept's functionality spans multiple different code modules but is experienced by the user as a single set of affordances.
    \item \textbf{Functional.} A concept is characterized by its functionality, and not how that functionality is visually presented in the user interface. So concepts will often correspond closely to backend services, although the mapping from concepts to services may not be one to one. The \textit{Clip} concept is rendered in a number of different ways, but retains a consistent identity regardless of its visual representation.
    \item \textbf{Behavioral.} A concept may sometimes be associated with an object-oriented class or an entity in a data model, but this is neither necessary nor sufficient. A concept embodies a dynamic behavior that typically involves multiple objects, and encapsulates its own data model. As an example, the \textit{Clip} concept can be refreshed from anywhere in the system, can reference a diversity of resource types (e.g. map sections or object views), and can be embedded into multiple resources. Understanding this holistic functionality is critical to using the \textit{Clip} concept effectively, even though users who use clips may be unfamiliar with the data structure of the \textit{Clip}.
    \item \textbf{Independent.} Unlike features, concepts are not defined with respect to some base functionality, but can be understood independently of other concepts, even if typically used in combination with them. The \textit{Clip} concept may be used with a \textit{Post} concept or some kind of \textit{Presentation} concept, but the user's mental model of clips is distinct and uncoupled from these.
    \item \textbf{Purposive.} A concept has a purpose that motivates its design, and addresses a particular user need. Having a purpose is key to assessing the value of a concept, and explaining to a user what the concept is for. The \textit{Clip} concept has a clear purpose of letting users embed segments of a given resource within another derivative resource while maintaining the linkage between the source and destination.
    \item \textbf{Reusable.} Most concepts can be applied in multiple contexts, across apps and within apps. Although concepts are implementation independent (so that a single concept can have multiple implementations), a single concept implementation can be reused in many contexts, so long as the implementation carefully preserves the independence of the concept (for example, by using type polymorphism appropriately). The \textit{Clip} concept provides a consistent experience across a heterogeneous set of sources (text, images, maps, videos) and destinations (documents, presentations, posts, emails).  
    \item \textbf{Valuable.} The concepts that are developed by a company over time as it builds and refines its products comprise the value that the company brings to the market. Concepts also embody the experiences and insights of developers not only in the behaviors and features they include but even in those they exclude. And because concepts are user-facing and evolve in response to user needs, the design knowledge associated with concepts includes lessons learned not only from the challenges of implementation but also from the users' environment in which the concepts are used. The \textit{Clip} concept provided value by synthesizing disparate unique individual insights from the many precursor concepts \textit{TextSnippet}, \textit{MapSlide}, and \textit{VideoClip} into a single concept and by strengthening the application suite by promoting cross-application connectivity.
\end{itemize}

\section{Developing the Concept Inventory} \label{developing_concept_inventory}
\begin{quote}
    The source of life which you create lies in the power of the language which you have---Christopher Alexander \cite{alexander_pattern}
\end{quote}
\begin{quote}
    Despite the efforts toward ``central planning,'' language (especially its everyday spoken form) stubbornly tends to go on its own rich, multivalent, colorful way.---James Scott \cite{scott_seeing_1999}
\end{quote}

\subsection{Implicit use of concepts in design work}\label{implicit_use_of_concepts}

Within \palantir, development activities fall into several categories: creating a new application that belongs to an ecosystem of existing applications; adding new functionality to existing applications; or consolidating the functionality of multiple existing applications into a single application or feature. Product managers, designers, and engineers engage with concepts in all of these situations, though traditionally the concepts have not been made explicit, and have not been treated as existing in their own right independently of the other concerns of the software at hand.

This conceptual work, until now only implicit, but nevertheless a critical part of all these categories of development, has primarily involved three distinct workflows:

\begin{itemize}
    \item During design, performing a “prior art” search to build context from older design mocks, understand the behavior of existing applications and features, explore the external landscape of production and experimental software outside of \palantir, synthesize documents written by colleagues in the past (often years and even a decade prior), and extract lessons and ideas for the design at hand;
    \item In response to reports of unexpected behaviors and UX problems, uncovering and resolving “concept bugs”;
    \item In evaluating new design proposals, building an intuition for how conceptually complicated the proposed software would be for end-users.
\end{itemize}

Prior-art searches serve the need to precisely understand the current state of a software system's concepts and functionality and the path it took to reach that state. They are archaeological explorations that at first take the form of a scratchpad of links, diagrams, quick notes, etc., and which then transition into a synthesis of the software system's concepts relevant to the goal at hand. Along the way, concept bugs emerge and are described in detail.

Examples of common concept bugs include concepts in different parts of a software system that share the same name but which have different meaning, concepts that share the same meaning but have different names, and unmotivated concepts that needlessly add complexity without providing the user any additional affordance. Additionally, the effort of precisely describing the present state of relevant concepts grounds the determination of how conceptually complicated a newly proposed feature or application will be.

At \palantir, although product managers, designers, and software engineers regularly wrote documents focusing on the concepts that comprise a piece of software, no shared language existed for structuring these documents and naming the concepts. Individual concepts existed, but the idea of a concept did not. The concepts, despite their centrality, did not exist as distinct entities apart from a specific implementation, and relationships between concepts and the features and applications that used them were not explicitly represented or tracked.

\subsection{Formalizing concept representations} \label{formalizing_concept_representations}
Our first, and most significant, step was to augment \palantir's internal ontology (i.e. the company-specific data model, backing store, and associated API endpoints for manipulating that data) to include concepts and relate them to existing entity types such as employees and software applications.

Internally, \palantir uses its \foundry software---software deployed to both commercial and government cust\-omers---to manage its ontology and run various business processes. Within that ontology, all software features, applications, services, capabilities, as well as products (including \foundry, \gotham, and \apollo) are represented as instances of a single \textit{platform component} entity type. Other entity types are linked to platform components, such as employees. For example, an employee might be a leader of a team of developers responsible for supporting the Object Storage v2 service.

\begin{figure}[ht]
  \centering
  \includegraphics[width=\linewidth]{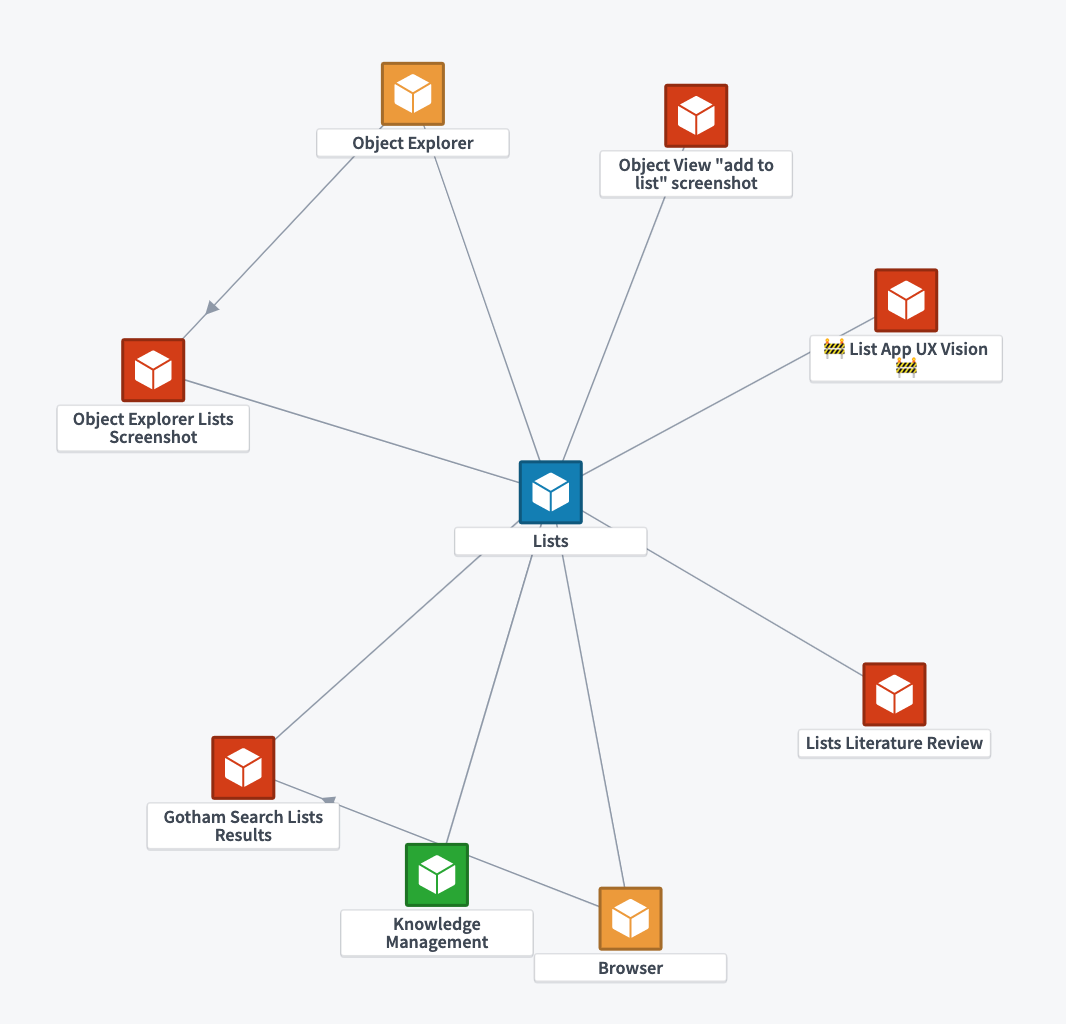}
  \caption{A screenshot of a graph of concepts, concept clusters, and resources as represented by \palantir's \foundry software system. The \textit{List} concept (blue, in the center) is referenced in various resources (red), including documents such as \textit{Lists Literature Review}, and screenshots of existing software such as \gotham \textit{Search Lists Results}, and in the \textit{Object Explorer} and \textit{Browser} platform components (orange), and it belongs to the \textit{Knowledge Management} cluster (green). 
  }
  \Description{A screenshot of a graph of concepts, concept clusters, and resources as represented by \palantir's \foundry software system.}
  \label{fig:concept_graph}
\end{figure}

Adding concepts to the ontology of entity types allowed us to formally represent both which concepts existed and which platform components used which concepts (see Fig. \ref{fig:concept_graph}). Furthermore, the concept entity type allowed external resources (including design mocks in Figma, prior art search and synthesis documents in word processors, and hand-drawn concept sketches) to reference and be referenced by the concept they illustrated. For example, we reified the concept of a \textit{List}, a “user-curated grouping of information intended to support decentralized knowledge management”.

EOS \cite{jackson_essence_2021} presents a different kind of concept graph that describes dependencies between concepts, in which the nodes are concepts alone, and an edge from a concept A to a concept B means that the inclusion of concept A only makes sense if concept B is included also. Given that we already had an extensive ontology of components, documents, features etc, it made more sense for us to augment our existing graph with concepts, showing the role they play with respect to these other entities. In future work, we plan to explore whether this concept dependence graph might be extracted from our graph, or whether adding it explicitly might be useful. 

\subsubsection{Concept Properties} \label{concept_properties}
Formalizing the representation of concepts in \palantir's internal ontology required deciding what properties ought to comprise a concept. We found that the most useful structure was simply to require two properties, a name and description. In EOS \cite{jackson_essence_2021}, concepts have a more elaborate structure (including, in addition to the name, a purpose, an operational principle, and state/actions). Motivated by feedback from users, especially product managers, we sought to simplify the data structure, lessen cognitive load, and enrich concepts by linking to external resources as needed. This had the added benefit of allowing for many people to contribute to a concept's contextualization over time without needing to achieve consensus about how to edit essential properties.

\subsubsection{Concept Aliases}
Standardizing a language of concepts requires people to learn new terms for existing ideas. No one person knows all the new terms, so a means of discovering what a familiar concept is now called is essential. Alternate names for concepts, which we called aliases, were our solution. They allow a user to search for ``Snippet'' or ``Snapshot'' (for example) and discover that what they're seeking is actually named ``Clip''.

Aliases may seem at first to be counterproductive. After all, having multiple names for the same concept may be confusing, and seems counter to the important role that names play for socializing patterns (whether in concept design or code). Aliases were introduced to support a transition from informal terms for concepts to a formal, organization-wide language. They allow for gradual adoption of canonical names for concepts by mapping multiple familiar terms to a canonical concept name, and allow external terms for the same concept to be incorporated.

\subsubsection{Concept Sketches} \label{concept_diagrams}
\begin{figure}[ht]
  \centering
  \includegraphics[width=\linewidth]{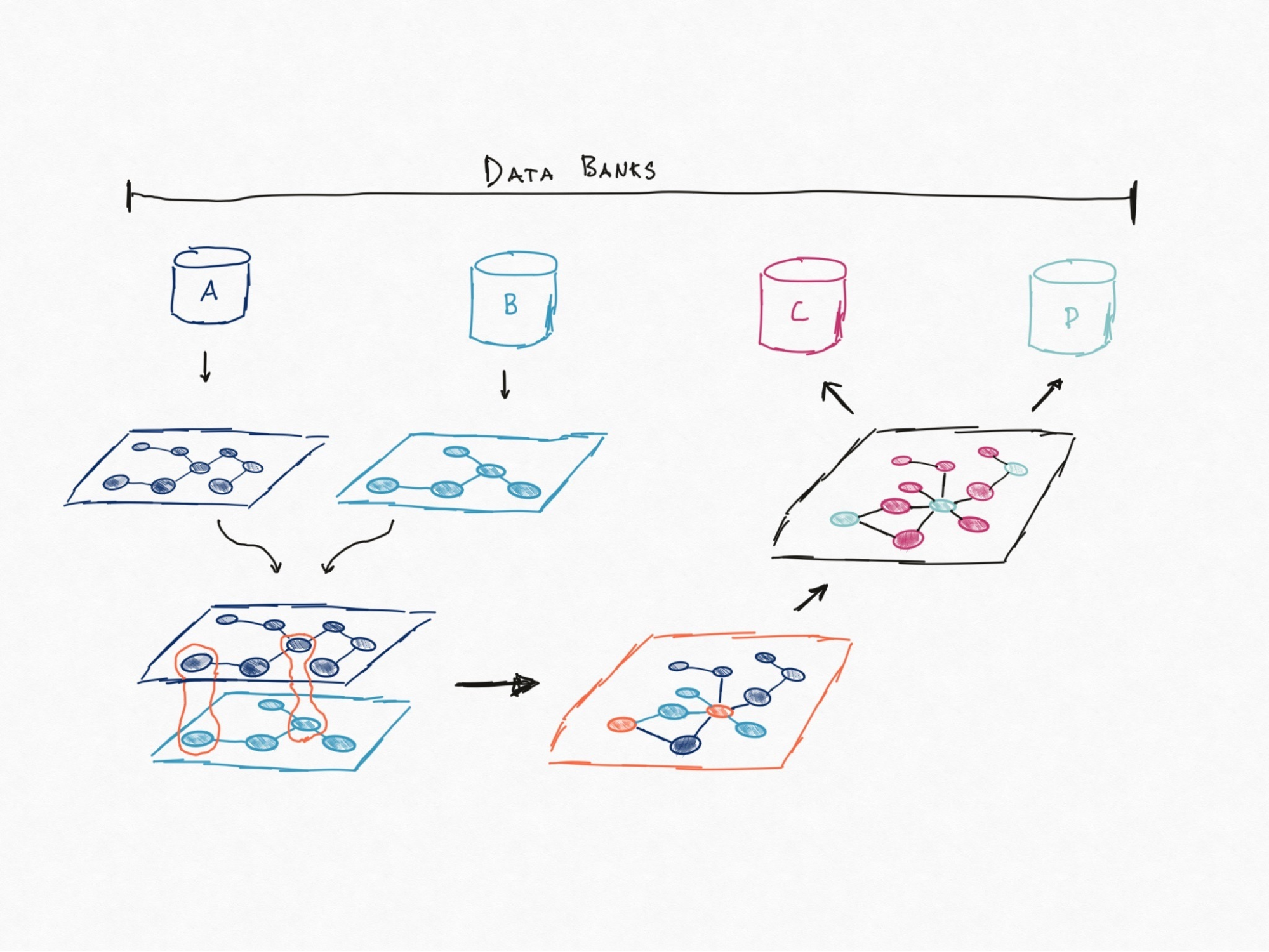}
  \caption{An example concept sketch for the \textit{Databank} concept, which is used by users building data pipelines that back entity types in an ontology.
  }
  \label{fig:databanks}
\end{figure}

Taking inspiration from Alexander's presentation of his pattern language \cite{alexander_pattern}, we embraced the concept sketch as a visual representation of a concept (see Fig. \ref{fig:databanks}). A sketch can convey ideas that would be laborious to explain with text, so leading with pictures helped make the concepts immediately accessible. We often draw sketches that bring several concepts together, allowing relationships between concepts to be shown visually.

\subsubsection{Concept Clusters} \label{concept_clusters}

We introduced a notion of \textit{concept clusters} to represent groups of concepts that are thematically related. Given that individual concepts can be quite granular, we needed a coherent but lightweight organizational layer above concepts that facilitated conversation about groups of concepts that frequently featured in planning, designs, feature requests, and other discrete work items.\footnote{The concepts in a concept cluster are typically used together; the clustering thus provides an approximation of the more detailed dependency graph of the kind advocated in EOS, which lets you define the dependencies of individual concepts, but is not symmetric, so that a dominant concept and an optional concept often used with it can be distinguished.}


Examples of concept clusters essential to product development work within \palantir include:
\begin{itemize}
    \item Files, Media and Attachments
    \item Data Entry and Tagging
    \item Ontology-Oriented Programming
    \item Collaboration
    \item Security
    \item Knowledge Management
    \item Search \& Discovery
    \item Object View Consistency
    \item Integrations, Federation and Writeback
    \item Data Transformation and Pipelining
\end{itemize}

Beyond a desire to simplify discourse about concepts, concept clusters were motivated by a desire to have specific product managers and designers own important cross-cutting concerns. Previously, product managers were responsible for an application or a group of applications; concept clusters were often owned only implicitly if they were owned at all. This was intuitive because applications have more or less clear boundaries, both in source code and in user interfaces, but it was deficient as an organizational structure because it failed to demarcate who was responsible for cross-cutting concepts used by many applications, capabilities, and features. As described above, \palantir's software products strive to offer a cohesive user experience across many different applications that make up a single product offering. Conceptual consistency is key to avoiding a fragmented user experience as users execute complex workflows that traverse multiple applications.

We included the notion of concept clusters so that a product manager could be made responsible not just for an application, but also for a concept cluster. The product development organization could now know who the responsible person was for a given collection of related concepts.

The reverse was also true: thanks to the relation between concept clusters and concepts (and between concepts and platform components), a concept cluster owner could easily understand which other product teams should be consulted when a concept or concept cluster needed to evolve to meet business demands. This was helpful especially when people shifted responsibilities---historically, people were implicitly associated with concepts long after they were explicitly responsible for them, and while this provided some continuity and consistency, it ended up with longer tenured employees accumulating concept baggage over the courses of their careers.

Concept clusters also afforded a new area of career growth for product managers, which we discuss later in Section \ref{assessing_impact}.

\section{Concepts in Action} \label{concepts_in_action}
\begin{quote}
    “Getting there (dynamics) is completely different than being there (statics). This is a distinction not only for academics but for practitioners as well.”---Hamilton Helmer \cite{helmer_7_2016}
\end{quote}

\begin{quote}
    “Philosophers have long wanted to understand concepts, but the point is to change them so as to make them serve our purposes better.”---Richard Rorty \cite{rorty_achieving_1999}
\end{quote}

\noindent Software development is an iterative process. Concept design is no different. Concepts are not static---they are subject to evolutionary pressure and the rate at which a given company can evolve its core concepts will often determine whether it succeeds or fails in the market.

As a higher-level construct, concepts evolve more slowly than the underlying source code. Because they are traditionally represented less formally than source code (if at all), disagreement about what concepts truly are and how they relate can fester invisibly, corroding the integrity of the product and eroding trust between teams. And because they are abstract and can evolve implicitly as code itself evolves, concepts can become distorted during the development process as microdecisions during the implementation process favor short-term feasibility over long-term value.

To remedy this, the process for evolving the concepts and concept clusters must be treated as an integral part of the development process. Changes to the concept inventory must be orchestrated with the same rigor as changes to the underlying source code in order to avoid persistent miscommunication. Just as with source code, ownership and review processes are critical aspects of an operationally relevant concept inventory. Changes need to be ratified by a central lexicographic authority; the current inventory needs to be easily queried by all members of the development team; major changes need to be disseminated proactively; and linguistic accuracy needs to be enforced through a decentralized culture of conceptual precision.

The development process must be extended so that it is always clear which concepts have been ratified but remain unbuilt and which concepts have been implemented and released. Because concept development generally precedes software development, valid concepts often lie dormant for many months, or even years, before their development is prioritized. This is especially true when introducing an entirely new concept cluster that may represent years of only partially parallelizable development effort.


\subsection{Concept Invention} \label{new_concepts}
Inventing a novel concept and delivering it to users is one of the true joys of product development. In our experience, the inspiration for a new concept often comes from one of two sources: the skeuomorphic incorporation of a physical artifact from the user's domain or an overarching system metaphor
(skeuomorphic innovation), or synthesizing two existing concepts into a third concept (dialectical innovation).

In all cases, though, good concepts serve a purpose---they are teleological, not just semantic, entities. A concept is always an invention. Even a concept that is not innovative was invented, by someone, at some earlier point. This is one respect in which the word “concept” can have inappropriate connotations because of its usage in other fields. Whereas for a philosopher the term refers to classifications or mental constructions that point to existing entities (the “concept of a dog”), for the software developer, concepts are hard-won inventions with real economic value \cite{romer}. The \textit{Folder} concept in today's file systems may seem to us now so natural that it's easy to forget the effort that went into creating and refining it (from its origins in early operating systems such as Multics to its modern variants in MacOS and Windows).


\subsubsection{Skeuomorphic Concept Invention} \label{skeuomorphic_concept_invention}
A major reason why inventing new concepts is risky is that they might be too foreign for users to grasp. For as long as programmers have created new interfaces, they have turned to skeuomorphism to bridge new concepts to what users already know.

\begin{figure}[ht]
  \centering
  \includegraphics[width=\linewidth]{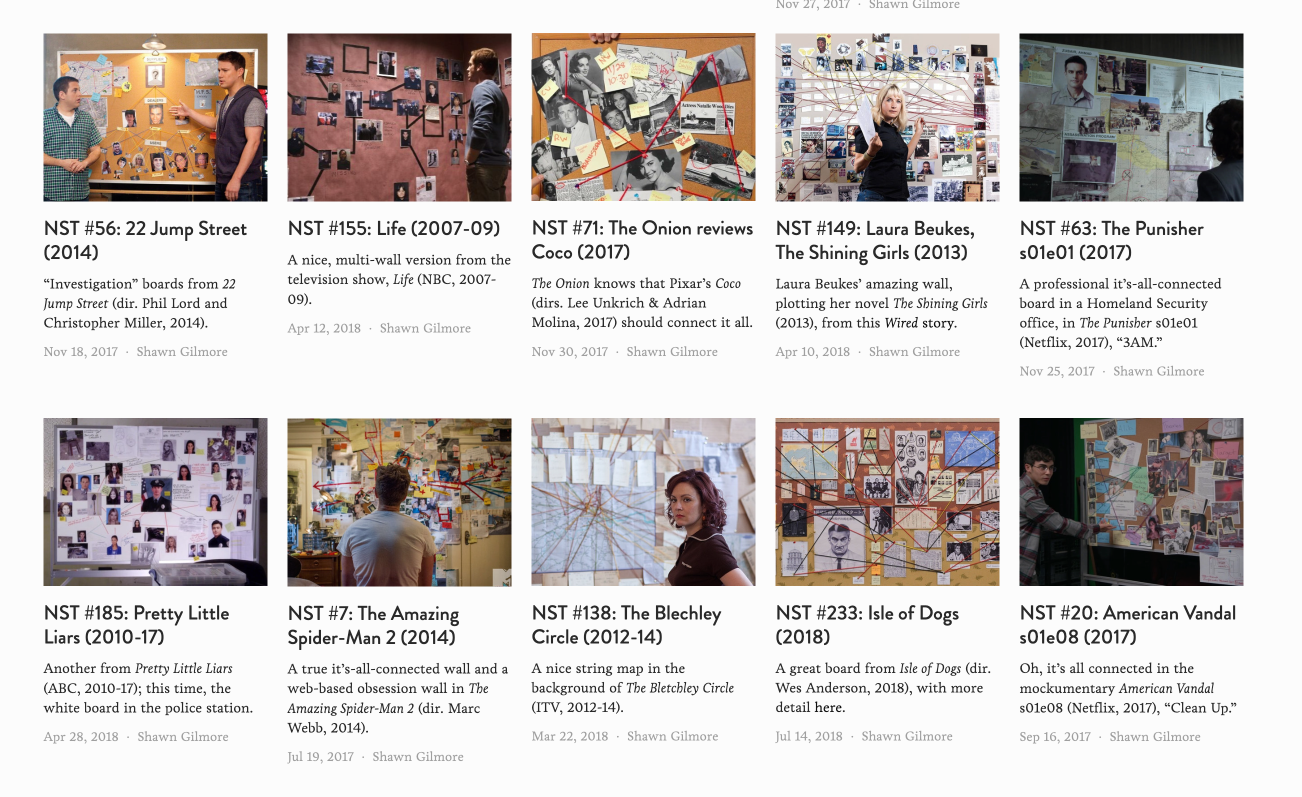}
  \caption{Almost every one of \palantir's early customers had one of these pinboards hanging in their meeting rooms. The ubiquity of pinboards, also known as evidence boards, is such that the general public is well-acquainted with them through movies and television as shown here in this collection of examples from The Vault of Culture publication. \cite{nst}}
  \Description{A collage of detective pinboards from TV shows.}
  \label{fig:pinboards}
\end{figure}

When the implementation of a concept resembles a thing that users are already familiar with, it's easier for users to reason about its affordances. 
At \palantir, for example, our ontology system was designed as a skeuomorphic parallel to the classic “detective's pinboard” (see Fig. \ref{fig:pinboards}). In contrast, the semantic web seems to have had no such parallel, perhaps explaining why (despite great interest in the early 2000s) the idea didn't gain traction in industry.\footnote{A concept may take on visual features of a familiar physical object, but have a different purpose and behavior (or have a similar purpose but a different visual representation). For example, the Macintosh \textit{Trash} concept adopted a familiar icon, helping users understand that they could delete files by moving them to the desktop trashcan. But the real purpose was not to allow \textit{deletion} of files but \textit{undeletion} \cite{jackson_essence_2021}, and the analogy to the physical trashcan offered no help with this. Similarly, while the UI of a \textit{ChatRoom} does not physically resemble a room of a building, it does help users understand how it separates different activities.}

\begin{figure}[t]
  \centering
  \includegraphics[width=\linewidth]{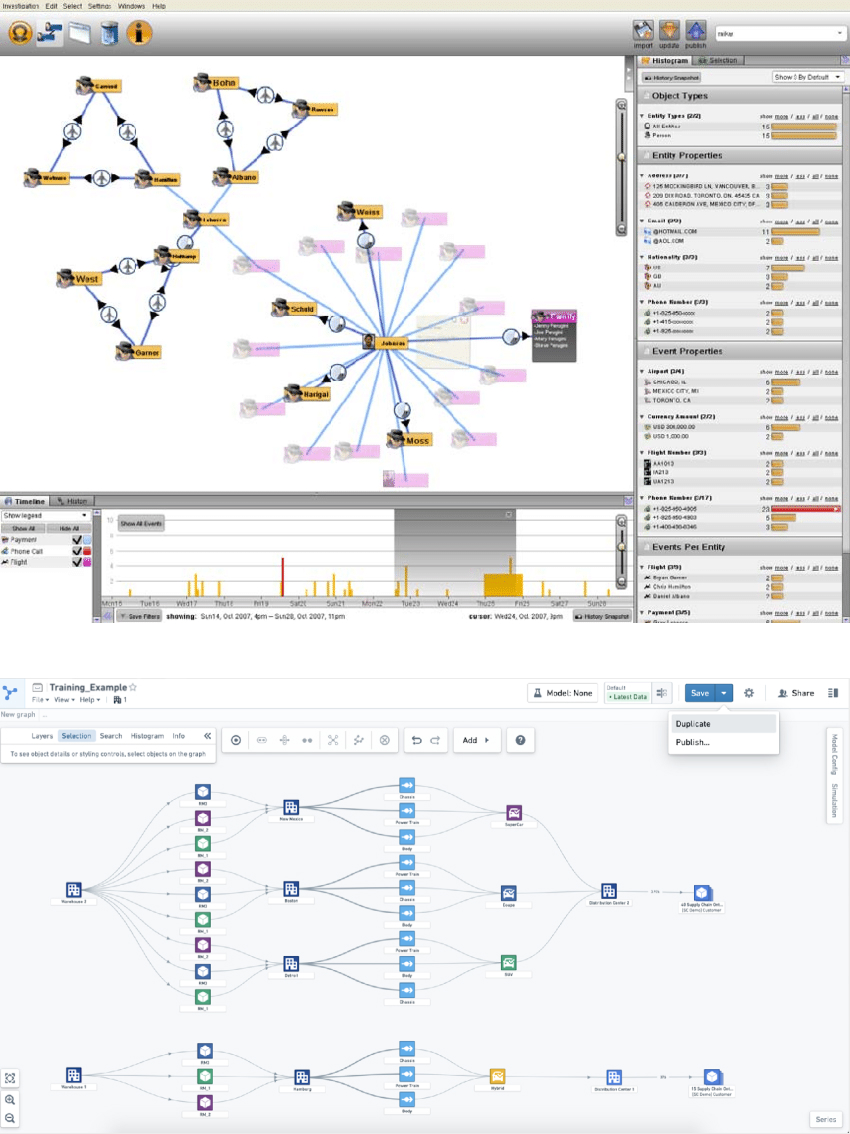}
  \caption{Top: an earlier version of \gotham's graph application relies on skeuomorphism to acquaint the user with key concepts, and helps them grasp the utility of non-skeuomorphic concepts (such a the interactive timeline at  bottom). Below: a more recent version of \foundry's graph application relies less on skeuomorphism since, during the time between \gotham and \foundry's launches, users had become more familiar with graph-based modes of interacting with data.}
  \label{fig:digital-pinboards}
  \squeezeup
  \squeezeup
  \squeezeup
  \squeezeup
  \squeezeup
  \squeezeup
  \squeezeup
  \squeezeup
  \squeezeup
\end{figure}

Our original \textit{Graph} concept was a digital representation of this physical analogy (see Fig. \ref{fig:digital-pinboards})---which allowed us to make progress on developing an ontology starting with the user experience, rather than the technology (following Steve Jobs's advice that user needs should set technology directions, and not vice versa \cite{jobs_wwdc_1997}). Our skeuomorphic approach grounded the system development, and also allowed us to introduce increasingly nuanced ontological concepts while anchoring them to existing workflows in the physical world.

Skeumorphism often applies to ontology design itself---the idea of a “digital twin,” so popular in industry, relies on the simple fact that representing the physical attributes of an organization is dramatically simpler than representing higher-level aspects like business processes and culture.


\subsubsection{Dialectical Concept Invention} \label{dialectical_concept_invention}
Concepts can also be invented as part of a classic dialectical process where new concepts are introduced in reaction to prior concepts---often synthesizing two previously oppositional concepts. This mechanism is often how the most novel concepts are developed. And, from a business perspective, concepts developed in this manner are often key sources of differentiation.

\begin{figure}
    \centering
    \includegraphics[width=\linewidth]{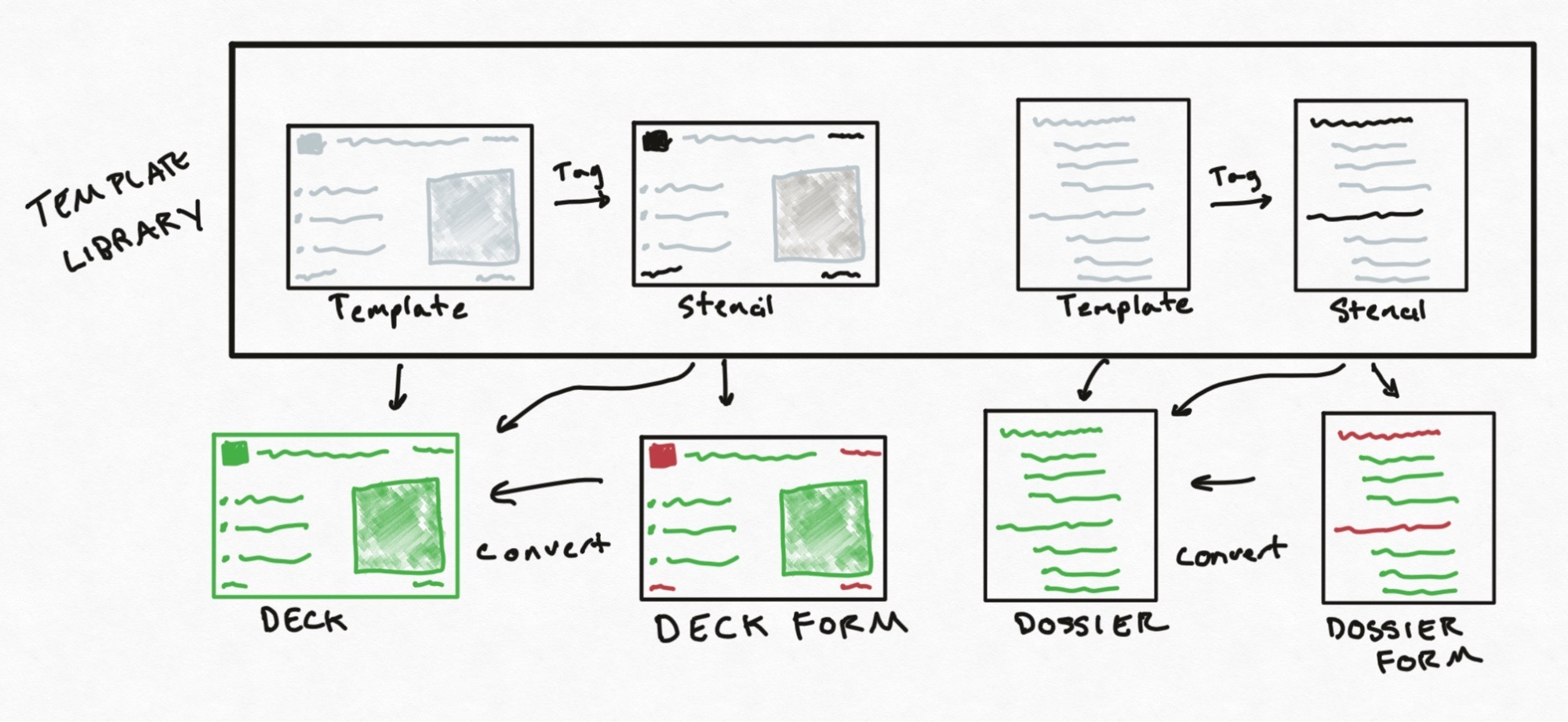}
    \caption{A concept sketch made during the process of creating the \textit{Stencil} concept, indicating the relationship between a \textit{Form}, a \textit{Template}, and a \textit{Template Library}.}
    \label{fig:templates}
\end{figure}

A concept that stands out as having been invented in this manner is what we call a \textit{Stencil}, which synthesizes elements of the \textit{Form} and \textit{Template} concepts (see Fig. \ref{fig:templates}).

Traditionally, forms are used for structured data entry, while canvas-based editors are used for unstructured document creation. Data created via forms is most commonly consumed using a tabular UI (e.g. as a Google Sheet) while documents created via a canvas-based editor are generally reviewed in the application which was used to create the document.

To enforce consistency, most canvas-based editors provide a concept of a "document template" which can be used to bootstrap a new document. Unlike forms, however, templates do not maintain their shape over the document lifecycle making them ill-suited for representing structured components of the document.

Inspired by PDF Forms, the \textit{Stencil} concept incorporated aspects of these two concepts by embedding form fields as elements within a document template and allowing these fields to be edited from either the form interface or canvas interface. As a result, document authors were able to bootstrap their document by filling out a form while still having the ability to create unstructured content within the document body while reviewers could consume data in both a document-oriented format (good for reviewing a single submission) as well as a tabular format (good for higher-level analytic workflows).

\begin{figure}
    \centering
    \includegraphics[width=\linewidth]{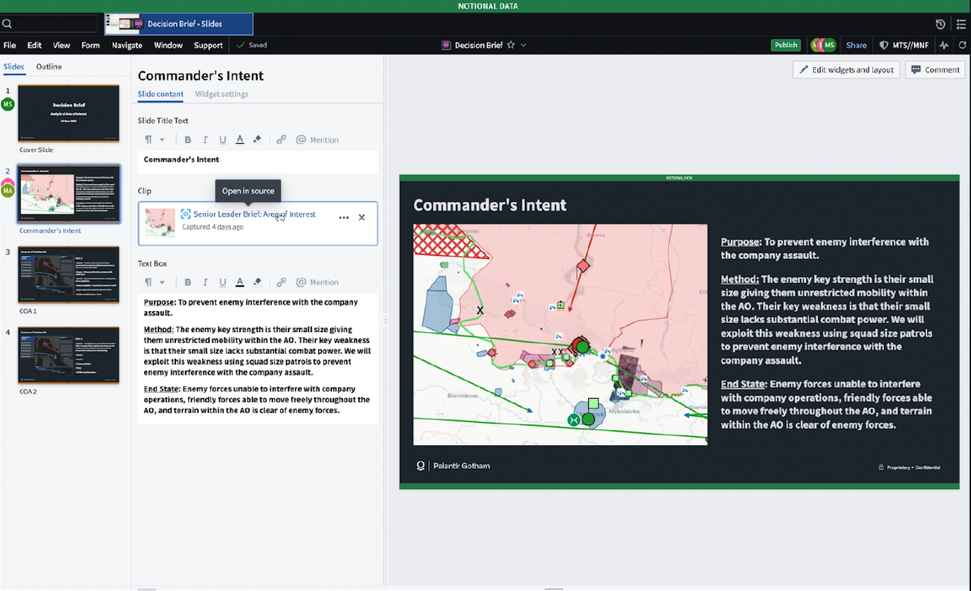}
    \caption{Stencils allow users to fill out a structured Form and produce a slide according to a Template.}
    \label{fig:slides}
\end{figure}

Using a stencil, just filling out a basic form autogenerates a correctly styled document, improving visual consistency across documents of a given type while also reducing the time users spent futzing with styling (see Fig. \ref{fig:slides}).

In addition, by synchronizing the \textit{Stencil} and \textit{Object} concept, producing templatized document exports for entities became as simple as mapping \textit{Object} properties to \textit{Form} fields; edits to these form fields could even be synchronized directly with the object, enabling novel ways to solicit object data edits from users and programatically generate templatized documents.

\subsection{Concept Refinement} \label{concept_refinement}
Over the course of the development of a product, its concepts evolve. Sometimes, a concept is refined or clarified; a single concept may also be split into a cluster of related concepts. More often, multiple concepts within a cluster evolve together, via some combination of additions, refinements and deprecations, with changes in one being balanced by changes in another.

In the appendix, we present in more detail two examples of concepts that evolved over time as product development teams realized the need to generalize application-specific concepts (\ref{case_study_clips}) and change multiple related concepts simultaneously (\ref{case_study_entity_resolution}).

\section{Assessing Impact} \label{assessing_impact}
\begin{quote}
``It doesn't matter whether a cat is black or white, as long as it catches mice.''---Deng Xiaoping \cite{vogel_deng_2013}
\end{quote}

\noindent Anyone who has worked for a large company will recognize the challenges involved in making significant changes to the development process. Indeed, we see the core contribution of our paper being the lessons we have learned (and the recommendations that follow) from this experience.

No matter how attractive a design theory may seem in the abstract, its ultimate test will be the quality of the results it produces: the experiences of users, the ease with which developers can extend and adapt functionality, and so on. 

It is too early for us to be able to offer any definitive conclusions about such impacts, but we can report on some promising early signs that our work is bringing benefits.

\subsection{Early Benefits} \label{benefits}
\subsubsection{Uptake of concept entities} \label{uptake_concept_entities}

\foundry tracks each entity's users, so we're able to watch the growth in the use of concepts as an entity type in the ontology. Back in January of this year, there were barely any ongoing users of the new concept entities, even though many had been added to the ontology. Now there are about 250 regular users within our development organization who read, write and query about 150 concepts. For comparison, the most widely used entity type is employee, which currently has about 1,500 users.

\subsubsection{Product Documentation} \label{product_documentation}
Organic adoption of concepts has percolated into product documentation, with many products maintaining an externally available user-facing concept inventory which is used to introduce the key concepts required to use a given application or product. As examples, both \palantir's \apollo DevOps platform and \foundry Time Series functionality document their concepts explicitly.

\subsubsection{Product manager career growth} \label{product_manager_career_growth}
One of the challenges for product managers as they progress in their careers is that they typically have few opportunities for promotion unless they disengage from product management work and oversee teams instead. We were hopeful that introducing concepts might address this problem, and allow product managers with deep technical expertise to be promoted while still contributing to design. We have already seen signs of this happening. One particular product manager, for example, has been given ownership of a cluster of concepts associated with cross-system functions such as authentication and network policies. This has given him the ability to influence multiple products and to play a major role simplifying the system integration experience. Prior to the introduction of concepts and concept clusters, this authority would have been informal, creating misalignment and creating friction in his work.

\subsubsection{Visibility for non-engineers} \label{visibility_for_non_eng}
In most organizations, roles are siloed, and communication between functions is limited. Consequently, design work that could benefit from the expertise of different roles becomes fragmented and is executed inconsistently with nobody having full ownership. Within \palantir, one particular challenge has been a lack of visibility into the engineered systems by those playing more client-facing roles. People writing proposals to clients  that would tout the benefits of particular capabilities of \palantir products had no easy way to access descriptions of those capabilities. It turns out that those capabilities are well aligned with product concepts. Now that the concepts at the heart of our products are clearly documented, along with their relationships to the products and the teams that can provide additional information about them, writers can more easily find the information they require, and then follow up with engineers if they need more.\footnote{When we passed our paper by our IP team for approval, they noted that their work too is often impacted by concepts having inconsistent nomenclature or shifting over time, and that a more conceptual view of functionality would help them determine when there are new issues to consider.}

\subsection{Adoption Challenges} \label{adoption_challenges}


The value of any inventory and its associated search engine depends on whether there's material worth searching for. The existence of years of documents, design mocks, and other resources created across \palantir was ready fodder for bootstrapping the corpus of concepts and clusters.

At first we sought the help of product managers (PMs) to translate existing documents into more formal concept and cluster descriptions. This failed for two reasons. Gathering resources and synthesizing a precise description of a concept or cluster takes time and effort. Although the entire organization benefits as a result, the author hardly benefits since they already are familiar with the ideas at hand: a classic collective action dilemma. Second, it was difficult to convey the utility of the project to someone if we couldn't show them a rich corpus of concepts, clusters, and their relations already in existence. We called this the “bootstrapping” problem.

The bootstrapping problem was straightforward to solve---we simply entered the first hundred concepts and dozen clusters ourselves. What we had assumed would take a long time merely required a few hours since we already had deep familiarity with a wide range of concepts, and we knew where historic and current documents and diagrams about those concepts resided.

The collective action problem of convincing others to volunteer their time to augment the concept inventory was harder, and, truthfully, is still a work in progress. Two efforts show signs of working. One is the alignment of organizational responsibility of PMs along the lines of clusters, which was discussed earlier in this paper. The other is building simple plugins or extensions for apps that PMs and other roles use every day. For example, we built an extension for the design software used across \palantir that lets users sync a concept sketch to \palantir's internal \foundry instance, along the way linking it to an existing concept. This dramatically reduces the time required to enrich the corpus of concepts and clusters. Since users don't have to leave the app they are already using, the cost of contributing falls. We prototyped a similar extension for a document editing application used at \palantir, which would allow a PM to sync documents to \foundry and link them to the appropriate concepts or clusters.


\section{Recommendations}

Our experience suggests some recommendations for other organizations that might want to deploy concepts at scale:

\begin{itemize}
\item \textbf{Inventory your concepts}. Set up a centralized inventory of concepts that can be easily accessed, searched and updated. At \palantir, we used our own software, but any shared document repository (eg, using Notion, Airtable, Google Docs, etc) should suffice if carefully structured and managed.
\item \textbf{Focus on user need}. Concepts should describe coherent, independent units of functionality, and should be driven primarily by user need (“ideas users must understand to use our software”), rather than by architectural concerns but may also include non-user-facing ideas essential to the development process.
\item \textbf{Record concept relationships}. Cluster concepts into groups, and record their associations with teams, products and code modules. Include informal materials in the inventory. Sketches in particular seem to help communicate concepts intuitively.
\item \textbf{Resist perfectionism}. Prefer inclusion in the inventory over correctness of individual concepts: better to have placeholders for concepts that have not been fully elaborated or clarified than to hide them.
\item \textbf{Invest in names}. Choose good concept names, and use them consistently in other documentation. Allow aliases to accommodate different names for the same concept, especially names used for the same concept externally.
\item \textbf{Focus on value}. Seed the concept inventory by focusing on breadth and what's pragmatically useful. Avoid adding concepts for the sake of completeness alone. 
\item \textbf{Teach your organization}. Create an accessible introduction to concepts that covers what concepts are, the role they play in meeting the organization’s needs, the practical value of curating a concept inventory, who will be responsible for concepts, and how the organization will recognize and incentivize concept ownership. Cultivate alignment among product managers, developers, and designers about how to communicate about concept bugs.
\item \textbf{Use concepts when planning}. Incorporate concepts into existing work planning and task tracking processes---achieve conceptual integrity via strategic integrity \cite{one_strategy}. Quarterly planning should reference priority concepts, and tickets should be linked to concepts when appropriate.
\item \textbf{Find concept owners}. Assign staff to the curation and maintenance of the concept inventory. Identify owners for concept clusters and treat concept development, refinement, deprecation, debugging, and maintenance as core responsibilities for product managers (or whichever role is appropriate for the particular organization).
\end{itemize}

\section{Background and Related Work} \label{background_related_work}

\begin{quote}
``I have consistently saved time and made better products by using BDUF (Big Design Upfront) and I'm proud to use it, no matter what the XP fanatics claim. They're just wrong on this point and I can't be any clearer than that.''
-- Joel Spolsky \cite{spolsky_2005}
\end{quote}

\subsection{Agility and Software Design} \label{agility_and_software_design}




\noindent A rush to agile methods has sometimes thrown the baby out with the bathwater \cite{meyer2014agile}. In our view, this is most apparent in the encouragement to start coding as soon as possible and to reject what is described derisively as ``big design upfront.'' Many software projects indeed suffered from creating elaborate design documents whose volume and detail were often not accompanied by similar levels of clarity and simplicity. Consequently, developers found themselves having either to implement needlessly complex functionality, or had to fill in holes in a vague design document without a full understanding of the system-wide consequences of their decisions. 

The solution to this problem, in our view, is not less design but better design. Our process places design at the center of development, but differs crucially from traditional processes in three key respects. First, our design artifacts focus on concepts (namely on abstract functionality), and do not detail every aspect of user interaction, which is often more effectively addressed in prototypes and in code itself. Second, our design artifacts are small and modular; we use concepts to embody the shared design assets of our products, and we do not construct monolithic design documents that contain all the design details of entire products. Third, our design artifacts are integrated into the development process, and evolve with it, acting as repositories of knowledge gained from the experience of users and developers in response to design decisions over time.

\subsection{Capability vs. Productivity} \label{capability_v_productivity}


Companies are often reluctant to invest in process improvements because of the short term costs, and may even cut back on capability investments, misreading a boost in productivity that follows for a positive sign, when in fact it is the harbinger of a longer term decline \cite{repenning2001nobody}. In software development, the notion of ``technical debt'' draws attention to the risks that short term expedients can incur.

In our effort, we sought a balance between the costs and risks of a  change in process (and potentially burdensome additional work) on the one hand and a capability investment that we believe will be of great benefit to our products and productivity in the future. This balance was achieved by gradual insinuation of concept design practices into our development workflows, and by using our ontology to integrate concepts into our existing documentation structures, leveraging the design documents we already had rather than requiring wholesale reworking.

\subsection{Objects, Roles, Views and Aspects} \label{objects_roles_views_and_aspects}

Concepts can be readily implemented in conventional languages. That said, concept structuring differs from object orientation and requires developers to adapt their perspectives. The key idea of OOP is to collect together in a single module all the behaviors associated with an object. With concept-centric design, in contrast, the same object identifier appears in multiple concepts, each governing a collection of behaviors that fulfill a particular purpose. A user's password, display name, and recent likes, for example, which might all belong to a single object in a classical object-oriented design, would likely appear in different concept modules: password in Authentication, display name in UserProfile, likes in Upvote.

In this respect, concept structuring is yet another approach---along with views \cite{jackson1995structuring}, roles \cite{reenskaug1996working}, subject-oriented programming \cite{harrison1993subject}, aspect-oriented programming \cite{kiczales1997aspect} and others---that critiques OOP for conflating the identity of an object with its involvement in behaviors, and seeks a different modularization for better separation of concerns \cite{dijkstra1982role}.

\subsection{Pattern Languages and Design Systems} \label{pattern_lang_and_design}

Christopher Alexander's idea of patterns \cite{alexander_pattern}, in which standard solutions to recurring problems are named and described, first influenced programming through the Gang of Four design patterns, which were archetypal solutions for certain problems (mostly related to removing coupling) in object-oriented code \cite{gamma1900design}.

A similar idea can be found at a higher level in architectural ``styles,'' in which software architectures can be classified into various known patterns \cite{shaw1996software}. By adhering to particular styles, developers are able to ensure some desirable functional properties, use tools more effectively, and avoid the complexity that arises from ad hoc structuring.

Software design patterns and architectural styles both focus on implementation structure. At the UX level, patterns are emerging under the rubric of ``design systems,'' which are collections of visual design elements along with guidelines or standards for how they are to be used. Design systems usually aim to bring more consistency to visual design across a company's products, or across applications that run on a particular platform. Design systems have their origin in Apple's Human Interface Guidelines (1978), but did not become popular until Yahoo's Design Pattern Library (2006), followed later by Google's Material Design (2014).

A design system typically includes a library of user interface elements, descriptions of common patterns of usage, and sometimes also graphical components for use in wireframing tools such as Figma. Design systems focus exclusively on the concrete user interface whereas concepts, like Alexander's work, address patterns in deeper functionality. They also tend to be focused on particular UI widgets and their proper usage, rather than being grounded in particular user needs. So the entry on progress indicators in Google's system, for example, does not offer the advice that users should be able to cancel a long operation (although Apple's guidelines do mention this); in the entry on date pickers, neither Apple's nor Google's guidelines address how to input a date range.

\subsection{Domain-driven Design}\label{domain_driven_design}

Like concept design, domain-driven design \cite{evans2004domain} seeks to orient the code around abstractions that are grounded in the world of the user. But whereas domain-driven design focuses on structures in the problem domain (in line with its precursors, most notably OMT \cite{rumbaugh1991object} and, earlier, JSD \cite{jackson1983system}), concept design recognizes recurrent structures in the functionality invented by designers to address user needs.

Domain-driven design suggests that ideas for software features and ways of implementing those features can be plucked from the fabric of a domain. In contrast, concept design allows for pure inventions without precedents in the domain. The boundary between discovery and invention is not always so clear, however, since a concept may have been invented in the physical world prior to computerization (as with the pinboard example of Section \ref{skeuomorphic_concept_invention}).

Domain-driven design's ``ubiquitous language'' plays a similar role to the language of concepts in providing a shared vocabulary and shared elements across a development team. But whereas domain-driven design is reluctant to encourage sharing across teams or products (with its notion of bounded context intending to provide firewalls between them), concept design highlights the sharing of key behavioral structures, in order to achieve alignment, consistency across products and reuse of expertise. The risk of introducing coupling between teams is addressed in concept design not by discouraging sharing, but by requiring that the shared concepts are independent of one another, and use types that are (unlike the entities of a domain) polymorphic and independent of any particular domain. In this way, concept design is helpful when building software that spans different domains. For example, the concept of Clips (\ref{case_study_clips}) is used across multiple Palantir applications that serve diverse user bases across multiple domains.

At Palantir, product development teams chiefly engage in concept design when building software. The work of customer-facing implementation teams that customize Palantir's software for customers' needs might be seen as more traditional domain-driven design, in which a custom ontology is crafted to match the customer's existing domain.

To recap, domain-driven design and concept design differ in two primary ways. First, whereas domain-driven design treats domain elements as \em{discovered}, concept design sees concepts as \em{invented}, either in the context of the development at hand\footnote{This corresponds closely to the distinction between "using ideas" and "producing ideas" - taken literally, the set of ideas accessible via domain-driven design will be bounded by the ideas already present within the domain \cite{romer}.}, or borrowed from previous developments. Second, concept design encourages modularity of software concepts and reuse across domains.

\subsection{Architecture and Organizational Structure}\label{architecture_and_organizational_structure}

Flaws in the structure of code have been recognized as sources of complexity and (what we now call) technical debt from the earliest days of programming. Coupling between modules, in particular, leads to knock-on effects, where a change in one module requires a compensating change to another, and so on. Much of the history of software engineering has involved strategies for identifying and reducing coupling. Parnas introduced the idea of representing coupling between modules as a dependency relation \cite{parnas1976design} (or equivalently, a graph or matrix \cite{eppinger2012design}), and formulated some principles of layered structure in terms of dependencies. Conway \cite{conway1968committees} suggested that the structure of a system and the organization that produced it would mirror each other (since the use of one module by another requires an agreement over the interface, and thus communication between developers); this claim has been explored further and empirically evaluated \cite{colfer2016mirroring}. In a related line of work, researchers have identified architectural antipatterns in the dependency graph that have been found to exact heavy costs \cite{mo2019architecture}.

But dependencies only capture explicit couplings. In a seminal paper, Parnas argued that modules should encapsulate design secrets \cite{parnas1972criteria}. The coding adage ``don't repeat yourself'' (DRY) is a simplified version of this idea. With only code to analyze, however, it is not easy to identify missed opportunities for modularization: the coupling between two pieces of code that do roughly the same thing is implicit, and cannot be identified without somehow mapping code elements to problem elements. Consequently, an easy (but usually bad) tactic to reduce apparent coupling is simply to replicate functionality.

Conceptual entropy is a kind of violation of DRY at a high level, in which similar functionality is repeated in different parts of an application and across applications, often in different variants. Eventually, it may be possible to infer opportunities for concept merging automatically from code (using LLMs, for example), but for now we believe that the first step should be making concepts and their relationships to code and teams explicit, as we have described.

\section{Conclusion}\label{conclusion}

Concepts are the underpinnings of any software system and the focus of many development activities. But they are often left implicit, and opportunities to simplify and align products are lost. Our experience suggests that by making concepts explicit in a shared inventory, product designs can be improved, siloing can be reduced, and entropic growth in software complexity can be curbed.



\bibliographystyle{ACM-Reference-Format}
\bibliography{for-arxiv}

\appendix
\section{Appendix: Concept Evolution Case Studies}

\subsubsection{Concept Evolution Case  1: Clips} \label{case_study_clips}
As with all design work, concepts need to be developed pragmatically---the best concepts are practical, working solutions to specific problems. But it's almost impossible for designers to anticipate all of the nuanced interactions and experiences that will surround a concept. As a result, concept clarification is a common workflow for concept designers.

\begin{figure}[ht]
  \centering
  \includegraphics[width=\linewidth]{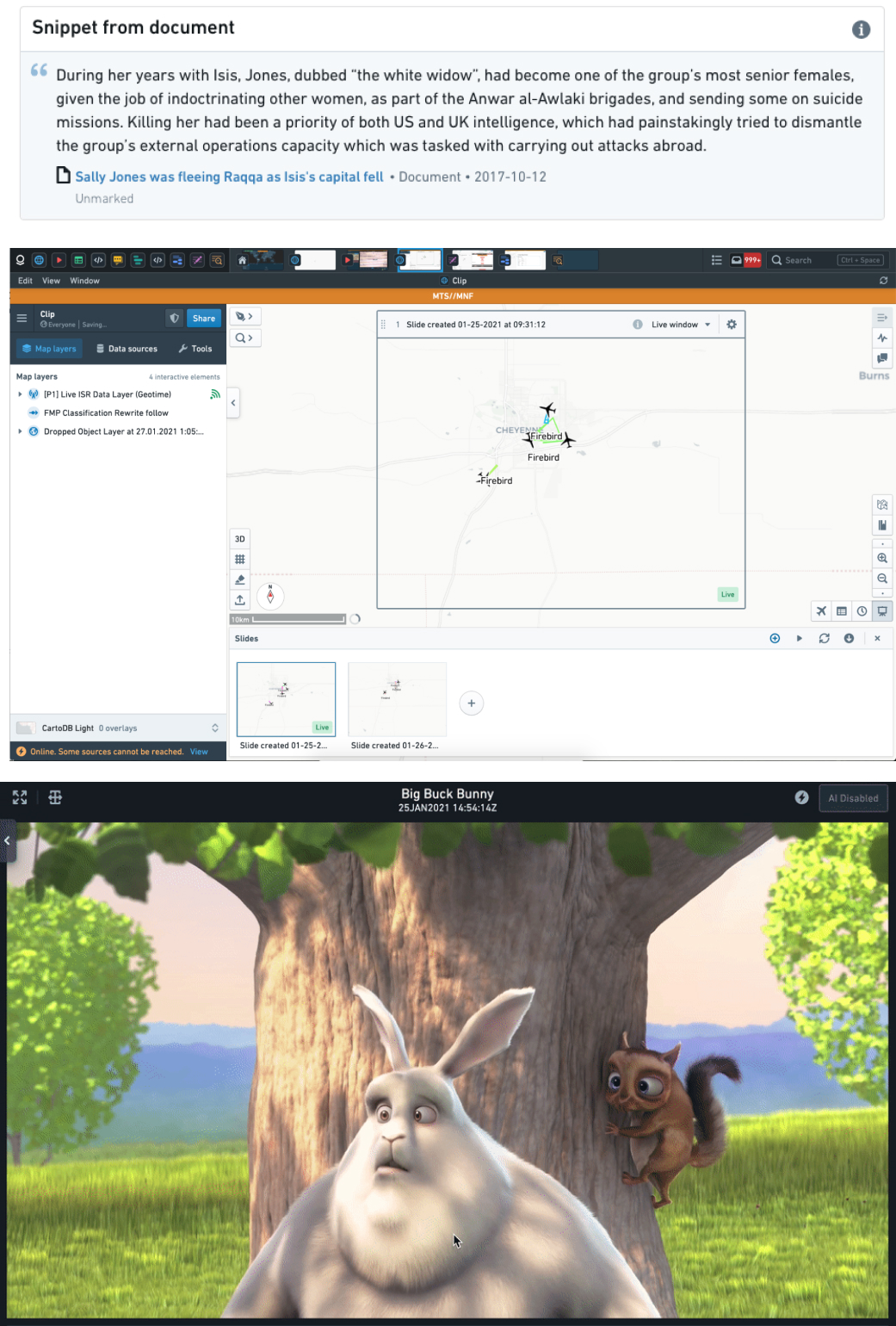}
  \caption{From top to bottom: an example of a \textit{TextSnippet}, a \textit{MapSlide}, and a \textit{VideoClip}, all of which existed as independent concepts prior to the generic \textit{Clip} concept.}
  \label{fig:snippets}
\end{figure}

In 2021, we began developing a generic \textit{Clip} concept to generalize a number of application-specific concepts: \textit{TextSnippet}, \textit{MapSlide}, \textit{GraphSnapshot} and \textit{VideoClip} (see Fig. \ref{fig:snippets}).

Inspired by concepts dating back to OpenDoc
in the 1990s, the purpose of clips was to let users embed segments of a given resource within another derivative resource while maintaining the linkage between the source and destination. Focusing on this specific purpose allowed us to generalize across both static and dynamic clip sources, by coupling an immutable “snapshot” of the underlying segment that had been clipped with the “metadata” required to regenerate a snapshot if the underlying resource changed. This allowed us to give clip consumers full control of their derivative work---something that most standard bookmarking and live preview concepts did not provide. If, for example, the source document was deleted, a derived document would still contain a valid reference.

After initial exploration in 2019, design started in earnest in January 2021 followed by development in mid-2021. By late 2021, the new concept had been released, and user feedback had started to accumulate. Throughout 2022, we monitored user feedback and started to synthesize it into some clarifications to the \textit{Clip} concept.

While the power of the concept was evident, users struggled to understand its version control semantics. In particular, as we sorted through the feedback, we realized that there were actually a number of versioning concepts at play---the version of the clip, the version of the resource that had been clipped, and the version of the clip that had been embedded within a document. This became particularly complex in the case of nested resources (resources that embedded other resources). When talking about the version of a clip, these concepts were being conflated.

\begin{figure}[t]
  \centering
  \includegraphics[width=\linewidth]{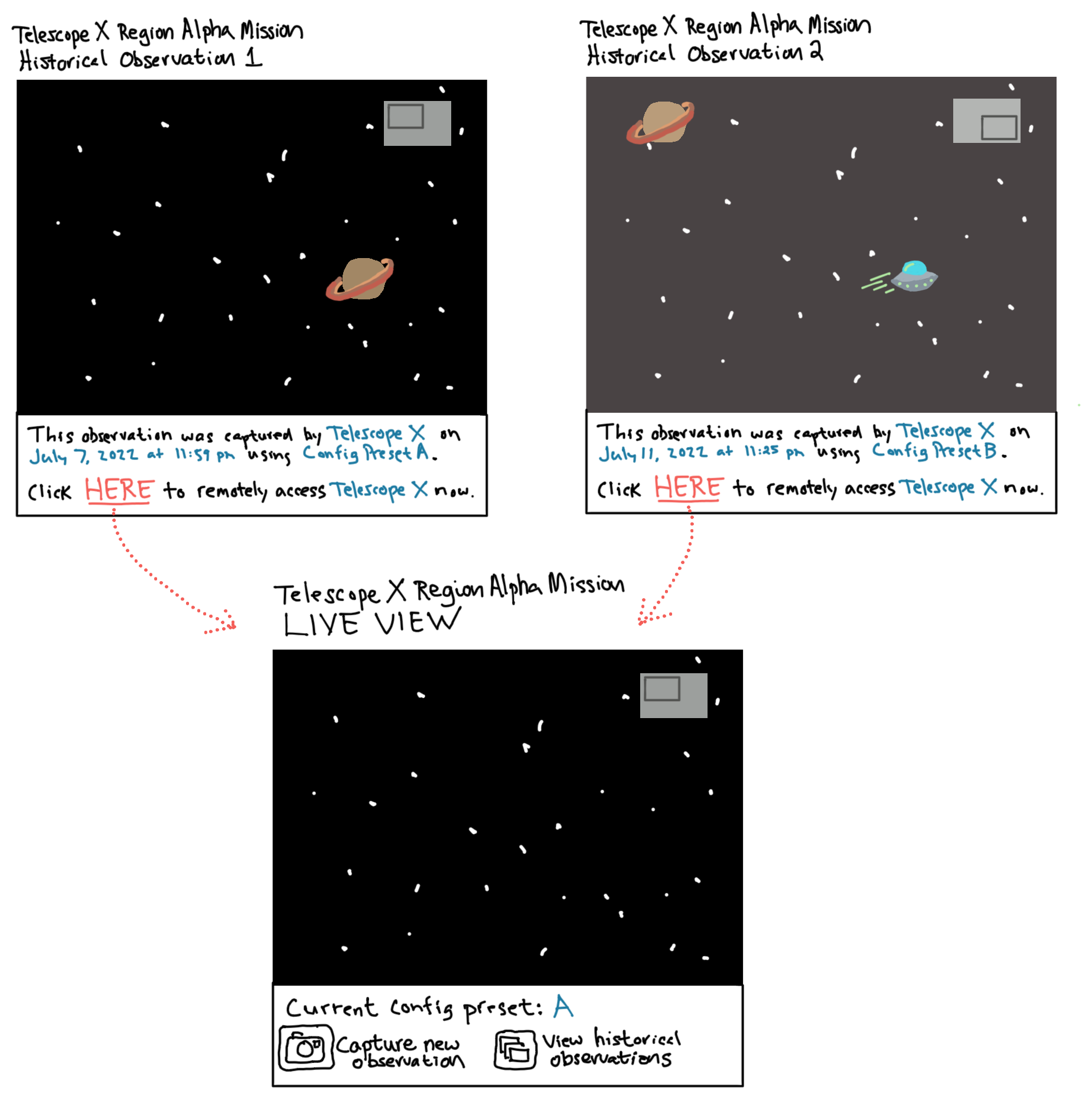}
  \caption{The team used the metaphor of an astronomer making observations through a telescope to explore the \textit{Clip} concept. Their description of this concept sketch reads:``Imagine an astronomer whose research project is to use a telescope to observe some special part of space. Whenever they see something interesting in that part of space, they record a new observation by saving what they see and when the image was captured, what settings they had on the telescope (the direction it was pointed in, etc).''}
  \Description{TBD}
  \label{fig:clips_diagram}
\end{figure}

To clarify the distinction between these different concepts, the team created an astronomy metaphor which helped them work through the different dimensions at play. Using this metaphor, the team drew a whimsical picture illustrating the various states that could exist, and how they might relate to one another (see Fig. \ref{fig:clips_diagram}).

Stepping back from the specific problem helped them think more abstractly about the concepts and with the help of the metaphor, the key challenge became much clearer---in some cases, users wanted to “change the telescope settings” and take a new picture, while in other cases, users wanted to record an updated picture with the same settings. In both cases, users wanted to be able to pivot between live and snapshotted views of the same region of space.

Using these insights, the team clarified the \textit{Clips} concept into a cluster of new, related concepts: \textit{ClipDefinition} to represent the clip metadata, and \textit{ClipCapture} to represent a static snapshot associated with a specific timestamp. New versions of a clip could be created by changing either the clip definition (which would create a new clip capture) or by manually refreshing the clip capture without changing the clip definition.

This split also clarified which affordances were available in a given location - clip captures could be regenerated from wherever the clip was embedded, whereas updating the clip definition required opening the clip in its source application.

\subsubsection{Concept Evolution Case 2: Entity Resolution} \label{case_study_entity_resolution}
Concepts evolve over time, in response not only to demands for new functionality and changes in a system's operating environment, but also as designers discover new and better ways to structure existing functionality. Typically, evolution occurs not just in a single concept but within a cluster, with several concepts changing simultaneously. In this section, we give an example of such an evolution.

In knowledge management systems (such as \palantir's), duplicate versions of the same real-world entity can exist for many valid reasons:

\begin{itemize}
    \item External systems can have duplicate entries for the same real-world entity (e.g. two CRM systems with duplicate customers).
    \item Data-as-a-service companies may ingest and normalize data from many primary sources into a derived data asset (e.g. Google Maps Places API).
    \item Due to syndication, a single system-of-record can be replicated in multiple forms in downstream systems (e.g. MLS data ingested into real-estate websites like Redfin and Zillow).
\end{itemize}

\textit{Entity resolution}, sometimes referred to as record linkage, is the process of combining these overlapping representations of the same entity to provide a merged view for the end user. It is a particularly tricky workflow. Poorly-designed entity resolution concepts can be inscrutable to end users, produce incorrect aggregations and false negative search results, and damage performance.

When rebuilding \palantir's entity resolution system in the mid-2010s, a critical requirement was solving not only the problem of resolution but also of \textit{unresolution}. Often, users would combine two entities, only for another user to come along and disagree. Because subsequent edits might have been made to the combined entity, unresolution exposed a wide array of subtle path-dependent errors influenced by the order of events in the system.

The first iterations of our design involved two key concepts. The \textit{ObjectContainer} concept was used to group properties together into objects with identities; the name of the concept reflects the fact that incoming data includes the values comprising the properties, but not always the objects themselves. The \textit{ResolutionGraph} concept was introduced to track the history of resolutions, with edges between objects showing when one object was merged with (and replaced by) another.

\begin{figure}[t]
  \centering
  \includegraphics[width=\linewidth]{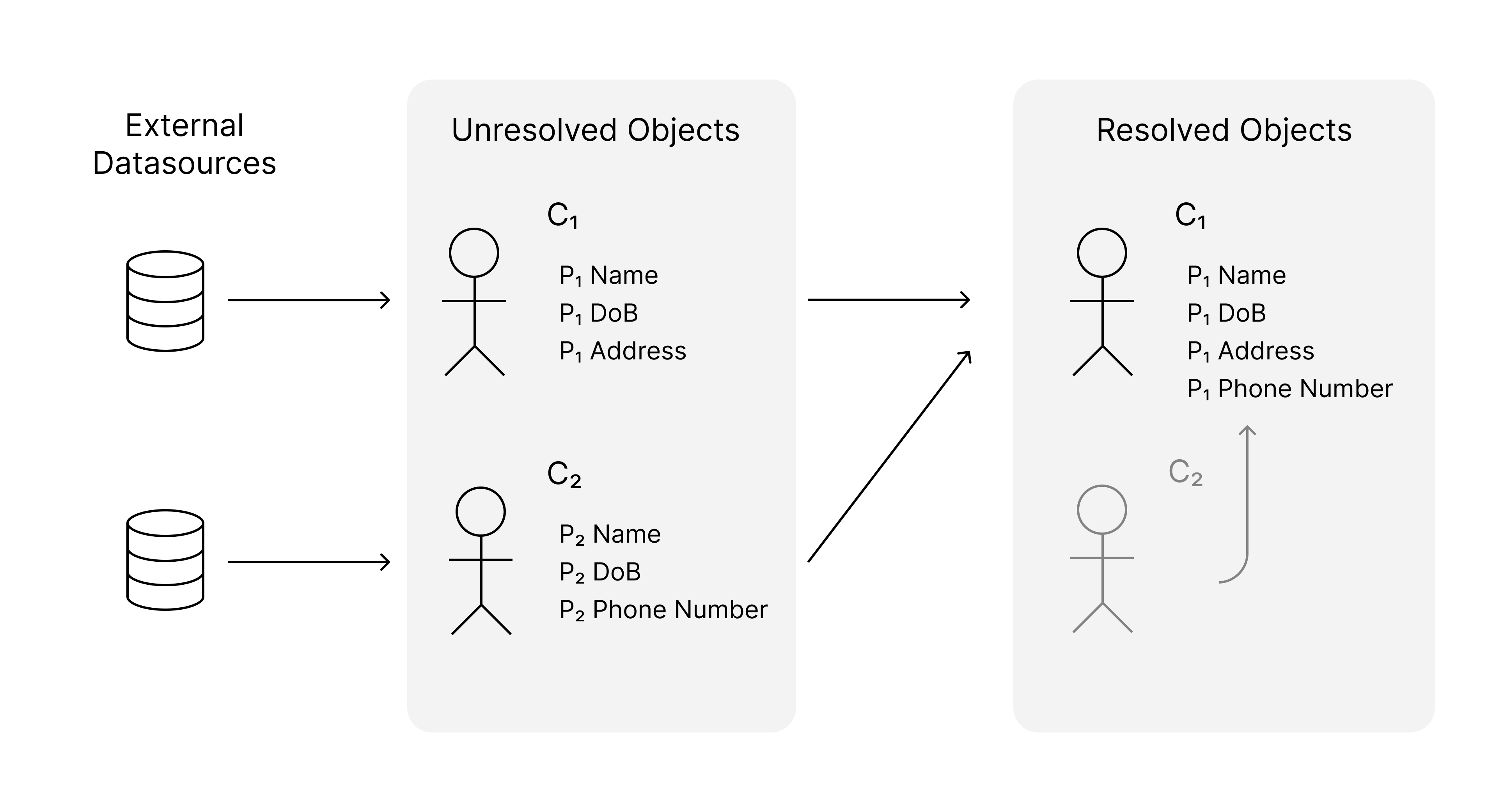}
  \caption{Resolution yields a single “winner” (C\textsubscript{1}) with all properties. Other Object Containers (C\textsubscript{2}) point to the winning Object Container. Property provenance is lost.}
  \label{fig:resolution_1}
\end{figure}

In this scheme, resolution involved moving the properties of one object container to another object container, turning the former into a kind of tombstone, and recording their relationships in the resolution graph (see Fig. \ref{fig:resolution_1}). 

While users appreciated the ability to resolve entities together, the resolution graph concept was poorly received by users, who found it unintuitive. Since the provenance of individual properties within an object container was lost, it was hard to perform unresolution completely reversibly.

Analyzing the concepts, we realized that our \textit{ObjectContainer} concept was overloaded---it was serving as both a grouping of properties, and a grouping for other object containers. In parallel, the \textit{ResolutionGraph} concept was attempting to perform bookkeeping for resolution history that overlapped with the role of the object container. In the degenerate case, where objects had not yet been resolved together, this was fine---there was no bookkeeping, and the contents of the object container were all uniform data values.

But as objects were resolved together, the system was unable to track which data values belonged to which original objects: data values were moved onto the \textit{winning object container} (the root node in the resolution graph), and needed to be segmented back onto the correct \textit{losing object containers} when objects were unresolved.

\begin{figure}[t]
  \centering
  \includegraphics[width=\linewidth]{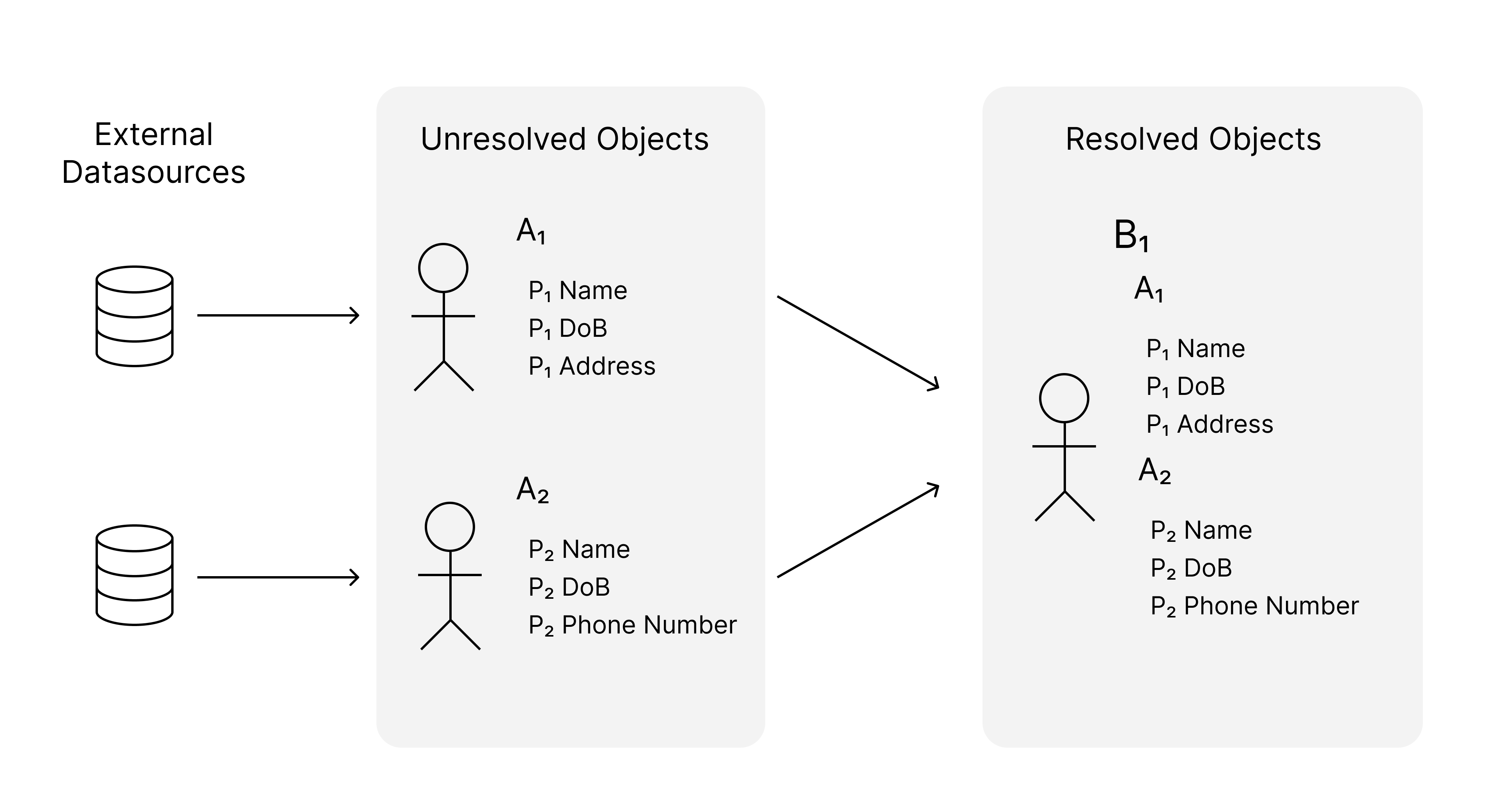}
  \caption{Resolution results in a single \textit{Bag} (B\textsubscript{1}) of \textit{Atoms} (A\textsubscript{1} and A\textsubscript{2}). Unresolution is possible because properties retain their connection to the \textit{Atoms} from which they came.}
  \label{fig:resolution_2}
\end{figure}

To solve these problems, we introduced two new concepts: the \textit{Atom}, which only contained properties, and the \textit{Bag}, which aggregated multiple atoms that had been subject to resolution. Now each concept was aligned more clearly with its purpose: atoms with structuring data, and bags with supporting resolution and unresolution. This radically simplified the unresolution process---users just needed to select which atoms to remove, and the system didn't need to use any complex heuristics to decide which data to bring along. The resolution graph concept was no longer needed; its role was subsumed by the new \textit{Bag} concept in combination with the existing \textit{EventLog} concept which maintained the history of events that led to particular bag structures (see Fig. \ref{fig:resolution_2}).

\end{document}